\documentclass[superscriptaddress,nofootinbib,amsmath,amssymb,prd]{revtex4-2}
\pdfoutput=1

\usepackage{slashed}
\usepackage{mathtools}
\usepackage{amsfonts}
\usepackage{amssymb}
\usepackage{epsfig}
\usepackage{empheq}
\usepackage{mathrsfs}
\usepackage{xcolor}
\usepackage{comment}
\usepackage{multirow}
\usepackage{rotating}
\usepackage[normalem]{ulem}

\usepackage{hyperref}
\usepackage{physics}
\usepackage{bm}

\begin{document}

\title{Testing cosmic acceleration from thermogravity without vacuum energy}

\author{Dong Ha Lee}
\email{dhlee1@sheffield.ac.uk}
\affiliation{School of Mathematical and Physical Sciences, University of Sheffield, UK}

\author{Jo\~{a}o Magueijo}
\email{magueijo@ic.ac.uk}
\affiliation{Abdus Salam Centre for Theoretical Physics, Imperial College London, Prince Consort Rd., London, SW7 2BZ, United Kingdom}

\author{Carsten van de Bruck}
\email{c.vandebruck@sheffield.ac.uk}
\affiliation{School of Mathematical and Physical Sciences, University of Sheffield, UK}

\author{Eleonora Di Valentino}
\email{e.divalentino@sheffield.ac.uk}
\affiliation{School of Mathematical and Physical Sciences, University of Sheffield, UK}

\date{}

\begin{abstract}
We present a first background-level observational test of a thermogravity theory in which the Einstein equations are clipped to a trace-free version, the cosmological constant $\Lambda$ does not gravitate, and controlled violations of energy conservation obstruct the usual reinstatement of $\Lambda$ as an integration constant.
We therefore set $\Lambda=0$ and ask whether late-time acceleration can instead be generated solely by energy non-conservation, with no extra background parameter relative to flat $\Lambda\mathrm{CDM}$.
We contrast a minimal universal implementation with a model in which only CDM partakes in non-conservation.
Using standard supernova distances together with uncalibrated BAO measurements, the universal model is strongly disfavoured, with the inclusion of DESI BAO worsening the fit by $\Delta\chi^2\simeq 51$ relative to $\Lambda\mathrm{CDM}$, because matter creation ties the intermediate-redshift normalization of $H(z)$ too rigidly to the present acceleration.
This conclusion, however, should be interpreted with caution, since universal non-conservation would modify the observational dictionary itself.
By contrast, restricting non-conservation to CDM leaves baryonic and photon observables unaffected at the background level and therefore allows a self-contained analysis.
The resulting $\xi\mathrm{CDM}$ model, which can be thought of as a one parameter extension of $\Lambda\mathrm{CDM}$, provides an improved fit relative to $\Lambda\mathrm{CDM}$ for the background dataset combinations considered, with improvements reaching $\Delta\chi^2=-4.85$ and a maximum Bayesian preference of $\ln\mathcal{B}=2.77$.
However, when the BAO ruler is calibrated using BBN or CMB information, the reduced effective early-time CDM density increases the sound horizon and drives $H_0$ towards lower values, thereby increasing rather than alleviating the tension with the distance-ladder calibration.
\end{abstract}

\maketitle

\section{Introduction}
\label{sec:introduction}

The standard $\Lambda$CDM model gives an economical and successful account of the observed Universe~\cite{Planck:2018vyg}, but its central ingredient remains obscure. The late-time acceleration is usually attributed to a cosmological constant $\Lambda$ or to some form of dark energy, while the vacuum energy expected from quantum fields is not observed to gravitate with its natural magnitude~\cite{Weinberg:1988cp,Padilla:2015aaa}. This is the infamous old cosmological constant problem. At the same time, the increasingly precise comparison between early- and late-Universe probes has exposed persistent tensions, most notably in the Hubble parameter $H_0$ and in the amplitude of late-time structure~\cite{Riess:2021jrx,DiValentino:2021izs,Abdalla:2022yfr,CosmoVerseNetwork:2025alb}. These tensions may eventually be blamed on systematic errors, but for the time being, they also motivate tests of alternatives in which the acceleration is not produced by a conventional dark-energy sector.

The model tested here arises from the marriage of General Relativity and thermodynamics. The starting point is the observation that black holes carry entropy and temperature~\cite{Bekenstein:1973ur,Hawking:1975vcx}, according to General Relativity and Quantum Field Theory. The logic, however, can be reversed: by applying thermodynamics to local horizon-like surfaces, Einstein's equations can be obtained as an equation of state~\cite{Jacobson:1995ab,Eling:2006aw,Jacobson:2015hqa}. 
We dub this framework {\it thermogravity}, using the term broadly for the thermodynamic and emergent-gravity perspective on gravitational dynamics~\cite{Padmanabhan:2009vy,Paranjape:2006ca,Verlinde:2010hp,Verlinde:2016toy,Chirco:2009dc}.
In the usual construction, the relevant thermodynamic process is purely a heat exchange. Gravity is then the macroscopic emergent manifestation of the system's response to heat flow across local null surfaces. This suggests a natural question: is General Relativity the whole thermodynamic theory, or only the degenerate limit of a larger one in which the microscopic system can also perform some form of ``work'' (for example, due to a chemical potential or any other extra thermodynamic dimension)?

The parent Letter to this paper~\cite{Isichei:2025ssf} answered this question by replacing the degenerate heat-only construction with an infinitesimal Otto-like cycle in a small causal diamond. The additional thermodynamic variable may be thought of as a coarse-grained number of microscopic space-time constituents, with a chemical-potential-like conjugate. The heat exchange typical of thermogravity still resides on the null boundaries of the diamond, but a volume-like work term arises, associated with the rest of the thermodynamic cycle. This immediately introduces a preferred local thermodynamic frame, the frame of the equator or waist (or, more physically, the mirror) defining the causal diamond. In the degenerate limit, this frame has no macroscopic effect. Away from that limit, it becomes physical, leading to controlled violations of local Lorentz invariance and of energy-momentum conservation.

Two dramatic consequences arise that are central to cosmology and its clash with the $\Lambda$ problem. First, the metric equations are clipped to their trace-free part. Vacuum energy is a pure trace contribution and therefore does not gravitate in the usual way. Second, the standard unimodular route back to General Relativity is blocked~\cite{Anderson:1971pn,Unruh:1988in,Henneaux:1989zc,Ng:1990xz,Smolin:2009ti,Padilla:2014yea}.
In trace-free gravity {\it with ordinary stress-energy-momentum conservation}, the missing trace can be reinstated as an integration constant, which is then interpreted as $\Lambda$. Here, however, the work term supplies a non-conservation law in the preferred frame, so the step that would restore an arbitrary cosmological constant is no longer available. The cosmological constant is therefore gravitationally inactive, and we ask whether the observed acceleration can be produced instead by the non-conservation processes of our theory.

At the homogeneous level, the preferred frame can be identified with the cosmological rest frame, and the mechanism has a simple interpretation: energy density is created, but momentum is not, in that frame. For dust, the minimal flat solution has the form $a(t)\propto t^{2/3}\exp(t/3L_2)$, so that the Universe behaves like the usual matter-dominated model at early times but accelerates at late times for positive $L_2$~\cite{Isichei:2025ssf}. The new scale $L_2$ is therefore not a dark-energy density in disguise. It is the time scale over which the thermodynamic cycle fails to be degenerate. In the background models considered below, this scale takes over the role usually played by $\Lambda$, without increasing the number of background parameters relative to flat $\Lambda$CDM.

The price is that this is not just another background expansion law to be fitted with the usual dictionary. If all species partake in non-conservation, then the same physics that changes $H(z)$ also changes the systems through which $H(z)$ is inferred. Redshifts, luminosity distances, baryon number, photon temperature, the baryon-to-photon ratio, stellar luminosities, and standard rulers may all acquire modified meanings. In such a universal model, it is therefore not self-consistent simply to import the standard supernova and BAO machinery unchanged. The universal case remains important because it provides an analytic benchmark and displays the mechanism in its cleanest form. But any exclusion obtained by applying the standard observational dictionary to it should be read as a formal exclusion of that naive implementation, not as a complete exclusion of universal thermogravity.

This motivates the second possibility studied in this paper. The non-conservation need not be universal. It may instead be confined to the dark sector, or dominantly absorbed by it. In the simplest such model, only cold dark matter is affected by non-conservation, while baryons, photons, and neutrinos follow their standard background evolution. This is less minimal from the microscopic point of view, but it is much cleaner observationally. Supernova redshifts and luminosities, BBN information, and the BAO ruler can then retain their usual operational meaning, at least at the background level. For this reason, the CDM-only model is the main self-contained target of the present analysis.

There is a further caveat. The present paper deliberately stops at the homogeneous background. This is not because perturbations are unimportant, but because in this theory they are unusually subtle. Once local Lorentz invariance and diffeomorphism invariance are broken, the preferred frame is physical. Choices that would be mere gauge choices in General Relativity may correspond here to different theories. Thus, the CMB, weak lensing, structure growth, and local gravitational tests cannot be obtained by blindly importing the standard gauge-invariant perturbation machinery of $\Lambda$CDM. They require a dedicated perturbative treatment of the preferred-frame model. The purpose of the present work is to find out which background implementations deserve that treatment.

The structure of the paper is as follows. Section~\ref{sec:model} summarizes the effective thermogravity equations used in the analysis and defines the universal and CDM-only implementations. Section~\ref{sec:background_cosmology} derives the corresponding background cosmologies. Section~\ref{sec:observations} describes the cosmological observations considered in the analysis, including Type Ia supernovae, BAO, BBN information, and compressed CMB constraints, together with the assumptions involved in applying them to the thermogravity models. Section~\ref{sec:results} presents the methodology, cosmological constraints, and model comparison for the universal and CDM-only scenarios. Finally, Section~\ref{sec:conclusions} summarizes our conclusions and discusses the limitations of the present background-level analysis and the need for a future perturbative treatment.


\section{The effective thermogravity model}
\label{sec:model}

We start by briefly summarizing the effective model derived in Ref.~\cite{Isichei:2025ssf}, without repeating the thermodynamic construction.
The starting point is a thermogravity derivation in which the usual heat-flow argument across local null surfaces~\cite{Jacobson:1995ab} is enlarged to a non-degenerate thermodynamic cycle.
In the macroscopic description, this has two effects. First, the metric field equations are trace-free, so that vacuum energy does not gravitate.
Second, the theory contains a preferred thermodynamic frame, so that the associated violation of local Lorentz invariance (and diffeomorphism invariance) implies that energy conservation need not hold in that frame.
The standard argument reinstating a cosmological constant (such as in unimodular-like theories) therefore fails.

There are several implementations\footnote{As explained in~\cite{Isichei:2025ssf}, the diamond structure used allows for the clipping of more than one Einstein equation, starting with the trace. Here we restrict ourselves to the minimal case.}, but in this paper we use the local trace-free implementation of the metric equations,
\begin{equation}\label{eq:tracefree}
R_{\mu\nu}-\frac{1}{4}R g_{\mu\nu}
=
8\pi G
\left(
T_{\mu\nu}-\frac{1}{4}T g_{\mu\nu}
\right).
\end{equation}
Since the trace of the metric equations has been removed, a cosmological constant appears only as an integration constant when energy is conserved.
Here we instead set the gravitating cosmological constant to zero and close the system by specifying a controlled violation of energy conservation, associated with the preferred thermodynamic frame of the theory.
The preferred frame is encoded in $n^\mu$, normalized by $n_\mu n^\mu=-1$, with spatial projector
$h_{\mu\nu}=g_{\mu\nu}+n_\mu n_\nu$.
For a single effective matter component, the projected non-conservation law used in Ref.~\cite{Isichei:2025ssf} is
\begin{equation}\label{eq:single nonconservation}
-n_\mu \nabla_\nu T^{\mu\nu}
=
\frac{1}{L_2}
\left(
n^\mu n^\nu+\frac{1}{3}h^{\mu\nu}
\right)T_{\mu\nu},
\end{equation}
where $L_2$ is the new length scale controlling the size of the effect.
The right-hand side is the light-cone average (as defined in~\cite{Isichei:2025ssf}) of the stress tensor in the frame selected by $n^\mu$.
For a perfect fluid comoving with this frame, it is simply $(\rho+p)/L_2$, so that positive $L_2$ corresponds to energy production in the homogeneous cosmological background.

The observational consequences depend crucially on how this violation is distributed among the different species.
We therefore use the multi-component generalization
\begin{equation}\label{eq:multi nonconservation}
-n_\mu \nabla_\nu T_I^{\mu\nu}
=
\left(
n^\mu n^\nu+\frac{1}{3}h^{\mu\nu}
\right)
\sum_J M_{IJ}T^J_{\mu\nu},
\end{equation}
where $I,J$ label the matter components.
The matrix $M_{IJ}$ specifies which sector fails to conserve energy and which sector sources the violation.
The minimal universal model corresponds to
\begin{equation}\label{eq:universal matrix}
M_{IJ}=\frac{1}{L_2}\delta_{IJ},
\end{equation}
so that every component obeys the same type of non-conservation law (and there is no cross-sourcing).
This model is useful as a benchmark, but it is not automatically a clean observational target: if photons and baryons participate in the violation, then the usual definitions of redshift, luminosity distance, baryon density, and standard rulers may themselves require rederivation.

The second model, and the main phenomenological focus of this paper, confines the violation to the cold dark matter sector.
In the simplest diagonal implementation,
\begin{equation}\label{eq:cdm matrix}
M_{cc}=\frac{1}{L_2}, \qquad M_{IJ}=0 \quad \mathrm{for}\quad (I,J)\ne(c,c).
\end{equation}
Baryons, photons, and neutrinos then obey standard conservation rules, while only the CDM density is sourced by the thermogravity non-conservation law.
This is the case in which the usual supernova, BBN, and BAO observables retain their standard operational meaning to the greatest extent, making the background analysis self-contained.

At the homogeneous level, we identify $n^\mu$ with the cosmological rest frame.
The background equations are then unambiguous and are derived in the next section.
Perturbations are more subtle.
Since local Lorentz invariance and diffeomorphism invariance are explicitly broken, different choices of the preferred frame beyond the background correspond to physically different theories rather than to gauge choices.
A proliferation of possibilities arises, with the resulting predictions being model-dependent and therefore non-generic.
For this reason, we restrict the present work to background observables and defer the perturbative treatment, including the CMB, weak lensing, and structure growth, to future work.

\section{Background cosmology}
\label{sec:background_cosmology}

We now specialize the effective equations to a homogeneous and isotropic background.
We first discuss the minimal universal model, for which all species obey the same non-conservation law.
This case is analytically useful and provides a benchmark.
We then turn to the CDM-only model, in which baryons, photons, and neutrinos obey their standard conservation equations.

\subsection{Universal violation}
\label{sec:universal background}

Consider a Friedmann-Robertson-Walker spacetime with scale factor $a(t)$ and spatial curvature $K$. In the universal model, each component obeys
\begin{equation}\label{eq:universal continuity}
\dot{\rho}_I+3H(\rho_I+p_I)
=
\frac{\rho_I+p_I}{L_2},
\end{equation}
where $H=\dot a/a$, and $L_2$ is the timescale associated with the non-conservation law.
The trace-free Einstein equations reduce to
\begin{equation}\label{eq:tracefree frw}
H^2+\frac{K}{a^2}-\frac{\ddot a}{a}
=
4\pi G\sum_I(\rho_I+p_I).
\end{equation}
For a pressureless component, Eq.~\eqref{eq:universal continuity} gives
\begin{equation}\label{eq:universal rho}
\rho_m(t)=\rho_{m,0}
\left(\frac{a(t)}{a_0}\right)^{-3}
\exp\left(\frac{t-t_0}{L_2}\right),
\end{equation}
where a subscript 0 denotes the present value.

The pure-matter, spatially flat case can be solved analytically.
Combining Eqs.~\eqref{eq:universal continuity} and \eqref{eq:tracefree frw}, one finds~\cite{Isichei:2025ssf}
\begin{equation}\label{eq:universal scale factor}
a(t)=a_0
\left(\frac{t}{t_0}\right)^{2/3}
\exp\left(\frac{t-t_0}{3L_2}\right),
\end{equation}
with
\begin{equation}\label{eq:rho constraint}
\rho_{m,0}=\frac{1}{6\pi Gt_0^2}.
\end{equation}
The corresponding Hubble rate is
\begin{equation}\label{eq:universal H(t)}
H(t)=\frac{2}{3t}+\frac{1}{3L_2}.
\end{equation}
Thus, the usual matter-dominated behaviour is recovered at early times, while for positive $L_2$ the solution tends asymptotically to accelerated expansion. There are other solutions to this theory, studied elsewhere~\cite{Dynathermo}, but we will focus on this one in this paper. 

It is useful to trade the integration constants $t_0$ and $L_2$ for quantities closer to those constrained by observations.
Defining
\begin{equation}\label{eq:Omega_m def}
\Omega_m=\frac{8\pi G\rho_{m,0}}{3H_0^2},
\end{equation}
Eq.~\eqref{eq:rho constraint} gives
\begin{equation}\label{eq:age to Omega_m}
t_0=\frac{2}{3H_0\sqrt{\Omega_m}},
\end{equation}
while Eq.~\eqref{eq:universal H(t)}, evaluated at the present time, gives
\begin{equation}\label{eq:L2 to H0}
L_2=\frac{1}{3H_0(1-\sqrt{\Omega_m})}.
\end{equation}
The present deceleration parameter is then
\begin{equation}\label{eq:universal q0}
q_0=-1+\frac{3}{2}\Omega_m.
\end{equation}
Current acceleration therefore requires $\Omega_m<2/3$.
More importantly for the fits below, Eq.~\eqref{eq:universal q0} displays the rigidity of the universal model.
Increasing $\Omega_m$ raises the intermediate-redshift normalization of the expansion history, but it also raises $q_0$, making the present acceleration weaker.
The model therefore has a built-in trade-off between the normalization of $H(z)$ at redshifts relevant for BAO and the low-redshift curvature measured by supernovae.

\begin{figure}
    \includegraphics[width=0.6\linewidth]{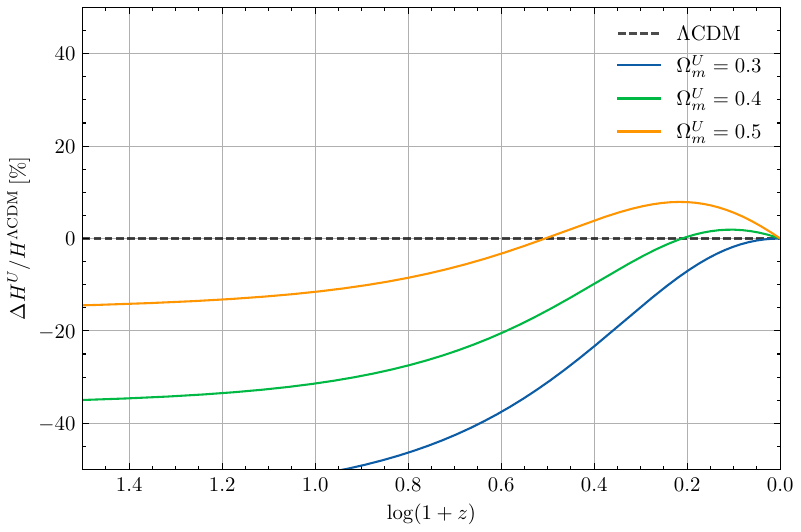}   
    \caption{Fractional difference between the Hubble parameter in the universal-violation model and in a fiducial flat $\Lambda$CDM model, $(H_U-H_{\Lambda{\rm CDM}})/H_{\Lambda{\rm CDM}}$, plotted at fixed $H_0=70\,{\rm km\,s^{-1}\,Mpc^{-1}}$. The $\Lambda$CDM comparison model has $\Omega_m=0.3$, while the universal-violation curves use the indicated values of $\Omega_m$. Increasing $\Omega_m$ improves the intermediate-redshift normalization of $H(z)$, but also raises $q_0=-1+3\Omega_m/2$, making the present acceleration weaker.}
    \label{fig:universal H fraction}
\end{figure}

For comparison with distance data, we write the solution in terms of redshift.
From Eq.~\eqref{eq:universal scale factor},
\begin{equation}\label{eq:universal z to t}
1+z=
\left(\frac{t}{t_0}\right)^{-2/3}
\exp\left(-\frac{t-t_0}{3L_2}\right).
\end{equation}
This can be inverted using the Lambert function,
\begin{equation}\label{eq:universal t to z}
t(z)=2L_2
W\left[
\frac{t_0}{2L_2}
\exp\left(\frac{t_0}{2L_2}\right)
(1+z)^{-3/2}
\right],
\end{equation}
which gives
\begin{equation}\label{eq:universal H(z)}
H(z)=
\frac{1}{3L_2}
\left[
1+
\frac{1}{
W\left[
\frac{t_0}{2L_2}
\exp\left(\frac{t_0}{2L_2}\right)
(1+z)^{-3/2}
\right]}
\right].
\end{equation}
This analytic expression will be used below as the background expansion history for the universal benchmark model.
Its qualitative behaviour is shown in Fig.~\ref{fig:universal H fraction}.
At fixed $H_0$, the universal model typically gives a lower $H(z)$ than the fiducial $\Lambda$CDM model over the redshift range most relevant to late-time distance probes.
This suppression can be reduced by increasing $\Omega_m$, but the same change also increases $q_0=-1+3\Omega_m/2$, thereby weakening the present acceleration.
The universal model is therefore too rigid: it tries to repair the intermediate-redshift normalization of the expansion history by moving in a direction that is disfavoured by the low-redshift curvature of the supernova Hubble diagram.

%
%

\subsection{Sourcing the CDM sector}
\label{subsec:cdm only}

We now consider the simplest non-universal implementation, in which the violation of energy conservation is confined to the cold dark matter (CDM) sector.
This is the cleanest late-time background model for observational purposes because baryons, photons, and neutrinos retain their standard conservation laws.
Thus, the usual interpretation of redshifts, supernova luminosities, BBN information, and BAO distances is not altered at the background level.
The CDM density obeys
\begin{equation}\label{eq:cdm continuity}
\dot{\rho}_c+3H\rho_c=\frac{\rho_c}{L_2},
\end{equation}
while all other species follow their standard background evolution.

Unlike the universal model, the CDM-only model does not lead to a closed analytic expression for $H(z)$.
We therefore integrate the background equations numerically.

It is convenient to use the number of e-folds,
\begin{equation}\label{eq:efolds}
N\equiv \ln a,
\end{equation}
with $a_0=1$ today.
Writing $E(N)\equiv H(N)/H_0$, the trace-free Friedmann equation becomes
\begin{equation}\label{eq:cdm background}
\frac{H_0^2}{2}\frac{dE^2}{dN}
=
K e^{-2N}
-
4\pi G
\sum_I\left(\rho_I+p_I\right).
\end{equation}
This equation, together with the CDM non-conservation law and the standard evolution of the other species, defines the background system used below.

For each species, we define the normalized density $f_I(N)\equiv\rho_I(N)/\rho_I(0)$.
Using the standard definitions of the density parameters, for $\Omega_K=0$ we obtain
\begin{equation}\label{eq:cdm ode1}
\frac{\mathrm{d}(E^2)}{\mathrm{d}N}
=
-3\Omega_b e^{-3N}
-4\Omega_r e^{-4N}
-3\Omega_c f_c(N),
\end{equation}
where the fractional CDM evolution is given by
\begin{equation}\label{eq:cdm ode2}
\frac{\mathrm{d}f_c}{\mathrm{d}N}
=
\left(\frac{\xi_c}{E(N)}-3\right)f_c(N),
\end{equation}
and we have defined the dimensionless parameter $\xi_c\equiv1/(H_0L_2)$, which quantifies the strength of the violation of energy conservation.

We may also take into account the small corrections due to massive neutrinos using the fitting formula
\begin{equation}\label{eq:neutrino evolution}
\rho_\nu(N)=\rho_{\gamma0}e^{-4N}
\left(\frac{7}{8}\right)
\left(\frac{4}{11}\right)^{4/3}
N_{\rm eff}
\left[1+(Ay)^p\right]^{1/p},
\end{equation}
where $A=0.3173$, $p=1.83$, and $y=m_\nu e^N/T_{\nu0}$, as described in~\cite{WMAP:2010qai}.

We solve these coupled equations numerically and convert to redshift using $N=-\ln(1+z)$.

As the dimensionless violation parameter $\xi_c\rightarrow0$, we recover the standard conservation equation for CDM, and the model becomes equivalent to standard $\Lambda$CDM at the background level, with $\Lambda$ arising as an integration constant.

\begin{figure}
    \centering
    \includegraphics[width=0.6\linewidth]{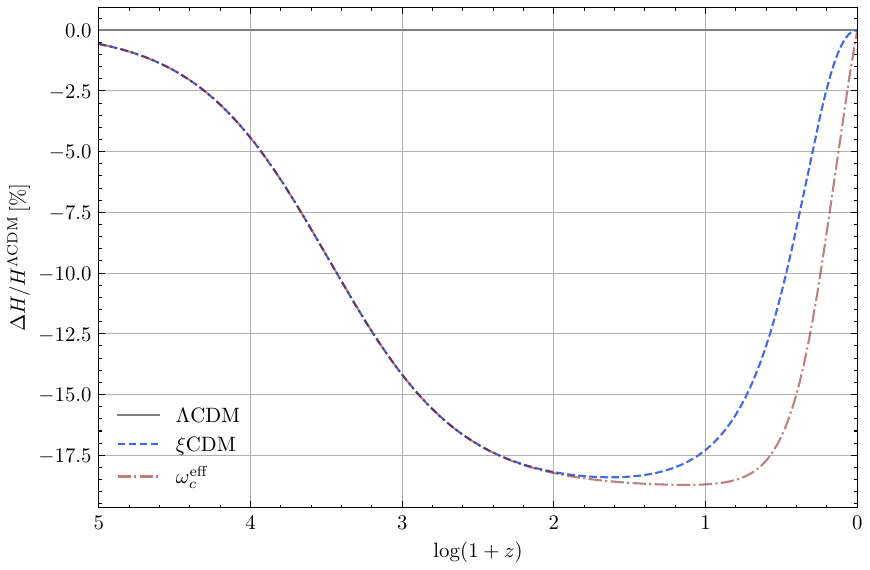}
    \caption{Difference in $H(z)$ with respect to a fiducial $\Lambda$CDM cosmology for the $\xi_c=0.5$ model sharing the same cosmological parameters (blue). The red curve shows the difference in $H(z)$ with respect to the same fiducial model for a $\Lambda$CDM cosmology with an effective CDM density $\omega_c\equiv\Omega_c h^2=\omega_c^{\rm eff}$, given by Eq.~\eqref{eq:effective cdm}.}
    \label{fig:cdm only H(z) residual}
\end{figure}

From Fig.~\ref{fig:cdm only H(z) residual}, we see that $\xi_c\neq0$ introduces a suppression in $H(z)$, relative to a $\Lambda$CDM cosmology, over the redshift interval probed by late-time distance measurements such as SNe and BAO.

\begin{figure}
    \centering
    \includegraphics[width=0.6\linewidth]{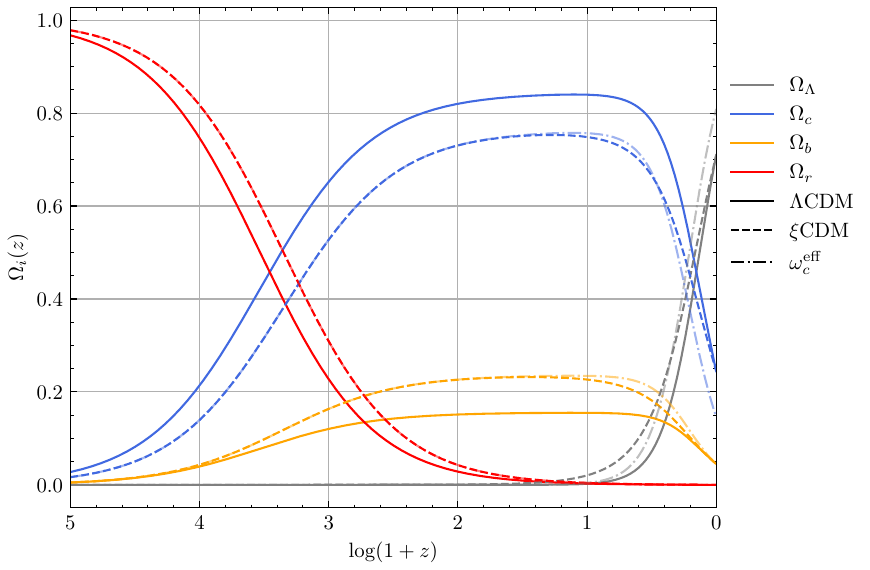}
    \caption{Relative energy densities $\Omega_i(z)\equiv\rho_i(z)/\rho_{\rm crit}(z)$ of the different components for $\Lambda$CDM (solid) and the CDM-violation model with $\xi_c=0.5$ (dashed). The fainter curves correspond to a $\Lambda$CDM cosmology with $\omega_c=\omega_c^{\rm eff}$ matched to the $\xi$CDM cosmology, as defined in Eq.~\eqref{eq:effective cdm}.}
    \label{fig:cdm only components}
\end{figure}

The physical effect of a non-zero $\xi_c$ is to reduce the CDM density in the past relative to a $\Lambda$CDM cosmology with the same present-day conditions, as shown in Fig.~\ref{fig:cdm only components}.
For $\xi_c=0.5$, this corresponds to a maximum decrease of $\sim40\%$ in the physical CDM density in the past.
Therefore, for more realistic cosmologies, we expect $\xi_c$ to be small (i.e. $L_2\gg H_0^{-1}$).

When one modifies the CDM sector in cosmology, additional care must be taken with observations that depend on the matter or CDM densities at a given epoch.
As will be discussed in further detail in Section~\ref{sec:observations}, when using BAO data or compressed CMB constraints to constrain a cosmological model at the background level, one often implicitly assumes standard CDM evolution to the relevant cosmological epochs (the baryon drag and recombination epochs, respectively) through fitting functions.

In other words, the physical CDM density today, $\omega_c\equiv\Omega_c h^2$, is taken as an input to the fitting formulae, assuming an $a^{-3}$ evolution to the relevant epochs.

To avoid implicitly assuming a $\Lambda$CDM cosmology when evaluating these quantities, we can use the analytic solution to the non-conservation equation for pressureless matter given by Eq.~\eqref{eq:universal rho}.
For $t\ll t_0$, where $t_0$ is the age of the Universe today, this allows us to define an effective present-day CDM density
\begin{equation}\label{eq:effective cdm}
\rho_c^{\rm eff}\equiv\rho_c e^{-t_0/L_2},
\end{equation}
which gives the same CDM density in the early Universe when evolved as $a^{-3}$.
We evaluate this quantity after solving the background evolution to obtain the present age of the Universe, $t_0$, and use it in the computation of the baryon drag and recombination epochs.

\begin{figure}
    \centering
    \includegraphics[width=0.6\linewidth]{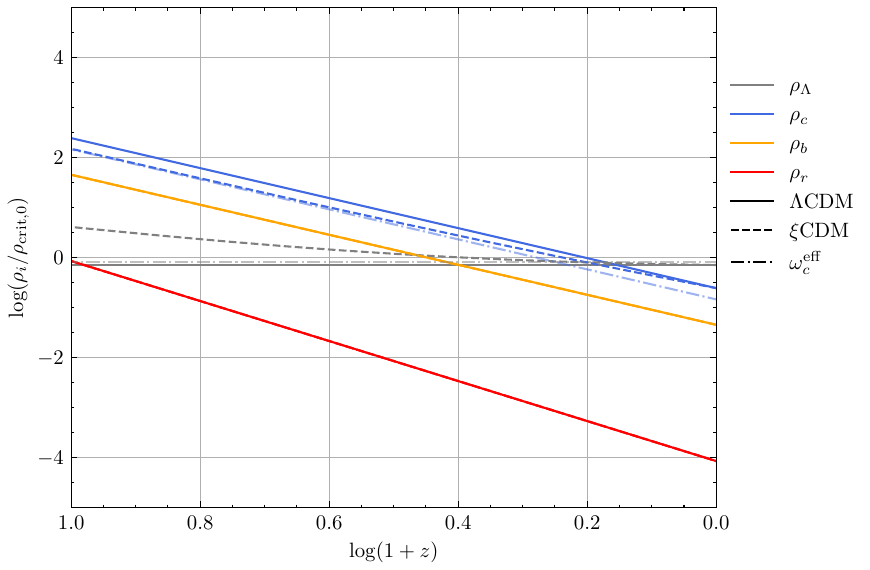}
    \caption{Evolution of the densities for $\Lambda$CDM, the $\xi$CDM model with $\xi_c=0.5$, and a $\Lambda$CDM cosmology with $\omega_c=\omega_c^{\rm eff}$. The CDM densities, shown in blue, illustrate that the early-time evolution of the $\xi$CDM model can be reproduced by a $\Lambda$CDM cosmology with the effective CDM density $\omega_c^{\rm eff}$.}
    \label{fig:eff cdm}
\end{figure}

We can observe from Fig.~\ref{fig:eff cdm} that Eq.~\eqref{eq:effective cdm} can be used at $z\gg1$ to effectively describe the physical CDM density at early times with an effective $\Lambda$CDM model.

\section{Cosmological Observations}
\label{sec:observations}

\subsection{Type Ia Supernovae}
\label{subsec:sn}

One method of probing the expansion history of the Universe is through measurements of cosmic distances to reference objects.
Type Ia supernovae (SNe Ia) are examples of such objects and act as standardizable candles, meaning that their magnitudes can be standardized to a fiducial absolute magnitude $M_b$.
The fiducial absolute magnitude sets the absolute scale of the background distances and is therefore completely degenerate with the dimensionful parameter $H_0$ of the cosmological model used to compute the distances to the SNe at their measured redshifts.
This degeneracy can be broken by calibrating the SN measurements with distance-ladder measurements, which provide the absolute scale of the SN luminosity based on astrophysical measurements and assumptions, independently of the cosmological model.

For a spatially flat cosmology, the luminosity distance $d_L(z)$ to a SN Ia measured at redshift $z$ is given by
\begin{equation}\label{eq:luminosity distance}
    d_L(z)=c(1+z)\int_0^z\frac{\mathrm{d}z'}{H(z')}.
\end{equation}
The distance modulus, defined as $\mu\equiv m-M_b$, can then be written as
\begin{equation}\label{eq:distance modulus}
    \mu(z)=5\log_{10}\left(\frac{d_L(z)}{\mathrm{Mpc}}\right)+25.
\end{equation}

In this work, we use the DES Year 5 SN catalogue with updated fits as presented in~\cite{DES:2025sig} (DES-Dovekie, DD), containing 1820 SNe in the redshift interval $[0.025,1.13]$, as our primary SN dataset.
We present the corresponding analysis using the Pantheon+ (PP) sample~\cite{Brout:2022vxf}, containing 1701 light curves of 1550 SNe in the redshift interval $[0.001,2.26]$, in Appendix~\ref{appendix} as a robustness check.

One can choose to calibrate the SN datasets using distance ladders such as that of the SH0ES collaboration~\cite{Riess:2021jrx} or the updated community-consensus distance-network measurement of $H_0=73.50\pm0.81\,\mathrm{km\,s^{-1}\,Mpc^{-1}}$ by the H0DN collaboration~\cite{H0DN:2025lyy}.
This allows us to break the $H_0-M_b$ degeneracy, provided that we assume the distance-ladder method to be robust.



\subsection{Baryon Acoustic Oscillations}
\label{subsec:bao}

Another method of probing the background cosmological evolution is through standardizable rulers.
Similar to standardizable candles, these are features that have a known characteristic length scale.

Baryon acoustic oscillations (BAO), produced by acoustic waves in the photon-baryon fluid in the early Universe, are imprinted on the correlation functions of tracers of the matter distribution and on the angular power spectra of the CMB.
Measurements of BAO from galaxy surveys such as the Dark Energy Spectroscopic Instrument (DESI)~\cite{DESI:2025zgx} can constrain the transverse comoving distance, which for a spatially flat cosmology is
\begin{equation}\label{eq:transverse bao}
    D_M(z)=c\int_0^z\frac{\mathrm{d}z'}{H(z')},
\end{equation}
and the line-of-sight comoving distance
\begin{equation}\label{eq:los bao}
    D_H(z)=\frac{c}{H(z)},
\end{equation}
relative to the comoving sound horizon at the baryon drag epoch, $r_{\rm drag}$.
The latter sets the characteristic scale of the BAO features imprinted in the matter distribution of the Universe.

In analogy with $M_b$ for SN distances, without knowledge of the calibrator $r_{\rm drag}$, we are not able to constrain $H_0$, which fixes the absolute scale of the cosmological expansion history.

The sound horizon at the drag epoch is given by
\begin{equation}\label{eq:r_drag}
    r_{\rm drag}=\int_{z_{\rm drag}}^\infty\frac{c_s(z)}{H(z)}\,\mathrm{d}z,
\end{equation}
where $c_s(z)$ is the sound speed in the photon-baryon fluid and $z_{\rm drag}\sim1060$ is the redshift of the baryon drag epoch, defined by the baryon drag optical depth reaching unity.

Therefore, unlike for SN distances, obtaining the absolute scale of the background evolution from BAO requires knowledge of the early-Universe physics of the cosmological model under consideration.

If left uncalibrated, BAO constrain combinations involving $H_0r_{\rm drag}$, relying only on the assumption that the BAO features observed in the matter distribution can be treated as standardizable rulers in the cosmology under consideration.

One method of determining the BAO scale is to use fitting formulae such as

\begin{equation}\label{eq:brieden r_drag}
    r_{\rm drag}\approx147.05
    \left(\frac{\omega_m}{0.1432}\right)^{-0.23}
    \left(\frac{N_{\rm eff}}{3.04}\right)^{-0.1}
    \left(\frac{\omega_b}{0.02236}\right)^{-0.13}
    \,\mathrm{Mpc},
\end{equation}
from~\cite{Brieden:2022heh}, calibrated around the Planck 2018~\cite{Planck:2018vyg} $\Lambda$CDM parameters.

These fitting formulae assume standard evolution histories for the species relevant to the computation of $r_{\rm drag}$.
This is typically not an issue for late-time modifications to $\Lambda$CDM in the dark energy (DE) sector, such as the $w_0w_a$CDM model, since at the epochs relevant to $r_{\rm drag}$, DE is negligible and the cosmology approaches the standard early-time evolution.

However, the models considered in this paper modify the evolution history of components that play a significant role at $z>10^3$ and therefore must be treated with some caution.

For the non-universal model described in Section~\ref{subsec:cdm only}, using the solution for $H(z)$ obtained by numerically solving Eq.~\eqref{eq:cdm ode1}, we can evaluate the integral in Eq.~\eqref{eq:r_drag} directly up to some $z_{\rm end}$ within the region of validity of our numerical solution, deep in the radiation-dominated (RD) era, and add the analytic tail of the integral in RD, where $c_s^2\rightarrow1/3$ and $H(z)\rightarrow H_0\sqrt{\Omega_r}(1+z)^2$.

The question then becomes how to determine $z_{\rm drag}$.
One may attempt to compute it numerically by treating the thermodynamics of the matter and radiation components in the early Universe, or resort to fitting formulae for this quantity, such as that described in~\cite{Aizpuru:2021vhd},
\begin{equation}\label{eq:aizpuru zdrag}
    z_{\rm drag}=
    \frac{1+428.169\omega_b^{0.256459}\omega_{bc}^{0.616388}
    +925.56\omega_{bc}^{0.751615}}
    {\omega_{bc}^{0.714129}}.
\end{equation}
Using a fitting formula for $z_{\rm drag}$ introduces a smaller error in $r_{\rm drag}$ than using a fitting formula for $r_{\rm drag}$ directly, particularly for extensions beyond $\Lambda$CDM, since in this case the approximation enters only through the lower limit of the integral, while the expansion history entering the integrand is computed directly for the cosmological model under consideration.

When modifying the evolution history of matter or radiation species in a cosmology with BAO, one must take into account that the parameters entering the computation of $r_{\rm drag}$ are present-day matter and radiation densities, which are then extrapolated to the relevant epochs assuming the standard $a^{-3}$ and $a^{-4}$ evolution for matter and radiation, respectively (unless the densities at each redshift are supplied directly to a recombination code).

Therefore, when using the fitting formulae described above, we must consider the effective present-day matter and radiation densities of a $\Lambda$CDM cosmology that would reproduce the early-time densities of the cosmological model under consideration.
These are the parameters that should be fed into the fitting formulae to ensure that the sound-horizon calculation correctly takes into account the deviations from standard $\Lambda$CDM evolution predicted by the cosmological model.

For the model corresponding to sourcing of the CDM sector, this is fairly straightforward.
Since the deviation from $\propto a^{-3}$ becomes negligible before the recombination epoch, we can obtain the effective CDM density in $\Lambda$CDM corresponding to the $\xi$CDM model, for a given value of $\xi_c$ and standard cosmological parameters, using Eq.~\eqref{eq:effective cdm}.

The analytic solution given by Eq.~\eqref{eq:universal H(z)} is more difficult to take into account.
The solution assumes that the violation of energy conservation is universal, but the analytic solution itself describes a matter-only Universe, with radiation negligible today.
This approximation is justified at the epochs where we measure SN and BAO ($z<5$), but not at the baryon drag epoch.
Furthermore, the universal violation directly affects the evolution of both matter and radiation, whose interplay is crucial for determining the BAO scale.
However, since the deviation is sourced by a single universal parameter $L_2$, the evolution of the different components is governed by the same underlying non-conservation mechanism.
We therefore assume that the relative abundances of matter and radiation at the relevant early epochs remain sufficiently close to their standard values that the microphysics determining the BAO scale is not significantly altered.
Under this assumption, the standard BAO ruler can still be used, although its calibration should be regarded with more caution than in the CDM-only model.
If we are able to consistently calibrate the BAO scale, we can combine BAO with other early-Universe probes to impose stronger constraints on the cosmological model.

By measuring the abundances of light elements in the Universe and assuming standard Big Bang nucleosynthesis (BBN), we can measure the ratio of the number densities of baryons to photons, $\eta\equiv n_b/n_\gamma$, at BBN.
Assuming that $\eta$ is conserved, we can obtain a measurement of the physical energy density of baryons today that is robust to late-time cosmologies, provided that the baryon mass remains constant.
This measurement of $\omega_b=0.02218\pm0.00055$~\cite{Schoneberg:2024ifp}, combined with a method for computing $r_{\rm drag}$, can act as an early-time anchor for BAO, analogous to how $M_b$, calibrated by the distance ladder, provides an absolute calibration of SN distances.
Knowledge of $r_{\rm drag}$ itself is only required if we wish to calibrate the absolute BAO scale, for example by combining BAO with an early-time probe such as the BBN measurement of $\omega_b$.
Otherwise, BAO can be used as an uncalibrated distance measurement, together with uncalibrated SN distances, still providing information about the shape of the expansion history.

\subsection{CMB}

For a full treatment of the Cosmic Microwave Background, we require computations at linear order in relativistic perturbation theory to obtain the power spectra predicted by the theory.
This is beyond the scope of this paper, but there are CMB observables from which we can extract additional constraints on the background evolution of the models.

A key observable in the CMB is the angular scale of the sound horizon at recombination,
\begin{equation}\label{eq:theta_s}
    \theta_s(z_{\rm rec}) = \frac{r_s(z_{\rm rec})}{D_M(z_{\rm rec})},
\end{equation}
where $r_s(z_{\rm rec})$ is the comoving sound horizon at the recombination epoch, $z_{\rm rec}\sim1090$, corresponding approximately to the maximum of the photon visibility function, and $D_M(z_{\rm rec})=(1+z_{\rm rec})D_A(z_{\rm rec})$ is the transverse comoving distance to the recombination surface.
The angular diameter distance is related to the luminosity distance by $D_A(z)=D_L(z)/(1+z)^2$, provided that the distance-duality relation holds.
This is closely related to the BAO scale described in Sec.~\ref{subsec:bao}, but is imprinted on the CMB photons rather than on the late-time matter distribution.
The two scales correspond to the different epochs $z_{\rm rec}$ and $z_{\rm drag}$, respectively.

However, it is important to keep in mind that CMB observations do not measure $\theta_s$ directly.
Instead, we measure the angular power spectra of the temperature and polarization anisotropies.
A value of $\theta_s$ inferred from the CMB therefore corresponds to that of the cosmological model which, at the level of linear perturbation theory, best fits the measured angular power spectra.

Under the assumption that the early-Universe physics of the model remains sufficiently close to that of $\Lambda$CDM, we may use the compressed CMB likelihood in \texttt{wgcosmo}\footnote{\url{https://github.com/williamgiare/wgcosmo/tree/main/likelihoods/CMB_compressed}}, which is constructed from the $\Lambda$CDM posterior obtained from Planck 2018 Plik high-$\ell$ TTTEEE + low-$\ell$ TT + low-$\ell$ EE, without lensing~\cite{Planck:2019nip}, and projects the posterior onto either the $2\times2$ or $3\times3$ compressed bases $\{\omega_b,\theta_s\}$ and $\{\omega_b,\omega_m,\theta_s\}$, respectively.

We should therefore interpret constraints involving compressed CMB likelihoods at the background level with caution and include them for completeness only.
For the $\xi$CDM cosmology, we choose to use the $\{\omega_b,\theta_s\}$ basis, since the baryon density follows the same evolution as in $\Lambda$CDM and provides an alternative early-time anchor for the BAO distances.

As discussed in Sec.~\ref{subsec:bao}, for the sound horizon, we can use a fitting function for
\begin{equation}\label{eq:aizpuru zrec}
    z_\mathrm{rec} = \frac{391.672\omega_{bc}^{-0.372296}+937.422\omega_b^{-0.97966}}{\omega_{bc}^{-0.0192951}\omega_b^{-0.93681}}+\omega_{bc}^{-0.731631},
\end{equation}
from~\cite{Aizpuru:2021vhd} and use the $H(z)$ obtained by numerically solving Eq.~\eqref{eq:cdm ode1}.

We choose to avoid the use of $\omega_m$, since the CDM density is directly modified by the model and there is therefore a greater risk of the constraints being influenced by the $\Lambda$CDM posteriors used to generate the covariance matrices for the compressed CMB likelihood.

Despite this, without a full perturbative treatment of the CMB, we cannot capture the full extent of the effects of the cosmological parameters of the model on the CMB.
We consider the $2\times2$ compressed CMB likelihood to extract as much information as possible from the observational tools available to us using only the background evolution, but the resulting CMB constraints should be interpreted with caution.

\section{Constraints from Observations}
\label{sec:results}

\subsection{Methodology}
\label{subsec:methodology}

The methodology of the analysis is as follows.
For a given cosmological model, we choose a set of cosmological parameters that together fully describe the model.
For this set of parameters, we impose our priors (typically flat over a wide interval for uninformative priors) and obtain constraints from a set of observational datasets described by their corresponding likelihood functions, using the Markov Chain Monte Carlo (MCMC) Metropolis-Hastings sampler implemented in \texttt{Cobaya}~\cite{Torrado:2020dgo}.
We consider the chains to be converged when the Gelman-Rubin statistic~\cite{Gelman:1992zz,Lewis:2013hha} satisfies $R-1<0.01$.

When comparing the performance of different models, we consider both the $\chi^2$ evaluated at the maximum a posteriori (MAP) point and the Bayes factor $\ln\mathcal{B}$ for each model and dataset combination.
We present
\begin{equation}\label{eq:delta chi2}
    \Delta\chi^2 = \chi^2_\mathrm{alt} - \chi^2_{\Lambda\mathrm{CDM}}
\end{equation}
with respect to $\Lambda\mathrm{CDM}$, such that $\Delta\chi^2<0$ indicates that the alternative model provides a better fit to the data.

For the Bayes factor, we use the learned harmonic mean estimator with normalizing flows method~\cite{mcewenMachineLearningAssisted2023,Polanska:2024arc}, implemented in \texttt{harmonic}\footnote{\url{https://github.com/astro-informatics/harmonic}} via \texttt{CosmoCAT}\footnote{\url{https://github.com/dlehdgk/CosmoCAT}}.
The Bayes factors are presented as
\begin{equation}\label{eq:ln bayes factor}
    \ln\mathcal{B} = \ln\mathcal{Z}_\mathrm{alt} - \ln\mathcal{Z}_{\Lambda\mathrm{CDM}},
\end{equation}
where $\mathcal{Z}$ denotes the Bayesian evidence.
With this convention, $\ln\mathcal{B}>0$ indicates that the data favor the alternative model over $\Lambda\mathrm{CDM}$ in terms of Bayesian evidence.

\subsection{Universal Thermogravity Model}

We impose a standard flat prior of $H_0\sim\mathcal{U}(40,100)$ km/s/Mpc.
For $\Omega_m$, since we require $\ddot{a}(t=t_0)>0$, we have $\Omega_m<2/3$.
Therefore, we impose a uniform prior of $\Omega_m\sim\mathcal{U}(0.1,0.67)$, where the lower bound is set in accordance with standard practice.
Since we use the simplified analytic solution of Eq.~\eqref{eq:tracefree frw} for pressureless matter, we acknowledge that Eq.~\eqref{eq:universal H(z)} is only valid during the late-time matter- and acceleration-dominated epochs.
Therefore, we choose to consider only combinations of the SNe ($z\sim1$) and BAO ($z\sim4$) datasets, without assuming any early-Universe calibration.
This requires us to drop any assumptions about the early-time physics determining the absolute scale of the BAO ruler, assuming only that the BAO signature exists in our Universal Thermogravity (UTG) model as a standardizable ruler.
Therefore, we choose to impose a flat prior of $r_{\rm drag}h\sim\mathcal{U}(90,110)$ Mpc, which is the quantity constrained by uncalibrated BAO.
We can, of course, add the distance-ladder constraints on $M_b$ (or equivalently $H_0$), which depend only on the geometric and astrophysical aspects of the distance-ladder calibration.
When we sample $\Lambda$CDM, we also choose to use the same priors to ensure that the prior volumes of both models are identical.

\begin{table*}[h]
\renewcommand{\arraystretch}{1.3}
\centering
\setlength{\tabcolsep}{8pt}
\begin{tabular}{l | c c c c}
\hline\hline
Datasets & DD & DD+H0DN & DD+DESI & DD+H0DN+DESI \\
\hline
$H_0\,[\mathrm{km}/\mathrm{s}/\mathrm{Mpc}]$ & $---$ & $73.51\pm0.82$ & $---$ & $73.51\pm 0.80$ \\
$\Omega_m$ & $0.418\pm0.016$ & $0.418\pm0.016$ & $0.5085\pm0.0073$ & $0.5081\pm0.0072$ \\
$r_\mathrm{drag}\,h\,[\mathrm{Mpc}]$ &  $---$ & $---$ & $95.80\pm0.58$ & $95.82\pm0.56$ \\
\hline
$\xi_c$ &  $1.061\pm0.036$ & $1.061\pm0.036$ & $0.861\pm0.015$ & $0.862\pm0.015$ \\
$L_2\,[10^3\,\mathrm{Mpc}]$ &  $4.33^{+0.60}_{-2.00}$ & $3.85^{+0.13}_{-0.15}$ & $5.32^{+1.00}_{-2.00}$ & $4.74\pm0.10$ \\
$t_0\,[10^3\,\mathrm{Mpc}]$ &  $4.73^{+0.60}_{-2.00}$ & $4.21\pm0.09$ & $4.28^{+0.90}_{-1.00}$ & $3.82\pm0.05$ \\
$r_\mathrm{drag}\,[\mathrm{Mpc}]$ &  $153^{+20}_{-50}$ & $136.2\pm7.9$ & $146^{+30}_{-50}$ & $130.4\pm1.6$ \\
\hline
$\Delta\chi^2$ & $-0.27$ & $-0.27$ & $51.4$ & $51.4$ \\
$\ln\mathcal{B}$ & $0.19$ & $0.16$ & $-25.85$ & $-25.82$\\
\hline\hline
\end{tabular}
\caption{$1\sigma$ constraints on the cosmological parameters of the UTG model for different dataset combinations using the SN dataset DES-Dovekie (DD). $\xi_c\equiv1/(H_0L_2)$ is a derived, dimensionless parameter quantifying the strength of the violations as presented in Section \ref{subsec:cdm only} describing the $\xi\mathrm{CDM}$ model.}
\label{tab:utg dd table}
\end{table*}

\begin{figure}
    \centering
    \includegraphics[width=\linewidth]{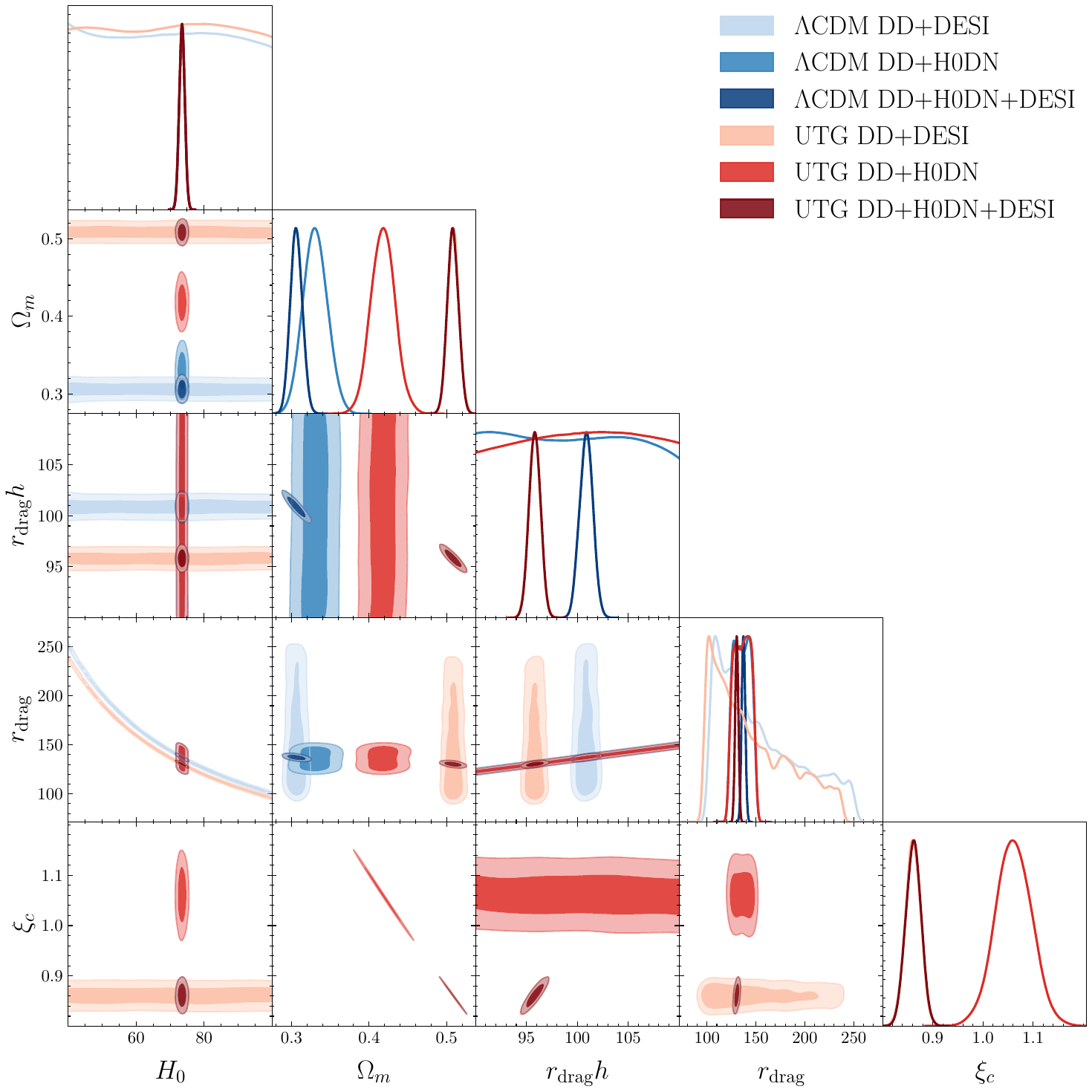}
    \caption{Posteriors of $\Lambda\mathrm{CDM}$ and the UTG model for the DES-Dovekie (DD) SN sample combined with other late-time probes.}
    \label{fig:utg triangle}
\end{figure}

From the constraints presented in Tab.~\ref{tab:utg dd table}, we observe that, in comparison to $\Lambda\mathrm{CDM}$ ($\Omega_m\sim0.3$), the UTG model appears to prefer a significantly higher matter density for all dataset combinations.
The higher matter density may be of interest in connection with the use of quasars as standard candles, for which, at higher $z$, the QSO luminosity distances appear to prefer higher values of the matter density, $\Omega_m\sim0.5$~\cite{Lusso:2025bhy}.
However, while the SN data alone can be well accommodated by UTG, the inclusion of BAO leads to a dramatic worsening of the fit, with a difference in $\chi^2$ of approximately $50$ with respect to $\Lambda\mathrm{CDM}$.
This indicates a strong incompatibility between the expansion history preferred by SN in UTG and that required to simultaneously fit the BAO distances.
We can observe from the posteriors shown in Fig.~\ref{fig:utg triangle} that, due to fundamental differences between $\Lambda\mathrm{CDM}$ and UTG, the two models occupy completely different regions of the parameter space for $\Omega_m$.
Once we consider the joint likelihoods of SN with either the distance ladder or BAO, the degeneracy directions involving $H_0$ and $r_\mathrm{drag}\,h$, respectively, are broken, and the two models remain confined to their distinct regions of parameter space.

We note that, as shown in Fig.~\ref{fig:universal H fraction}, the UTG model intrinsically has a slower expansion rate in the past relative to $\Lambda\mathrm{CDM}$.
This, combined with the rigidity of the model, which has only two cosmological parameters ($H_0,\,\Omega_m$), the same as $\Lambda\mathrm{CDM}$, means that it fails to jointly fit the SN and BAO distances measured at different redshifts.

\subsection{$\xi\mathrm{CDM}$ model}
\label{subsec:cdm model}

We impose the same standard priors for $\{H_0,\omega_c,\omega_b\}$ and a flat prior $\xi_c\sim\mathcal{U}(0,1)$, where the limit $\xi_c\rightarrow0$ recovers $\Lambda\mathrm{CDM}$.
Since we are able to compute $r_\mathrm{drag}$ using $\omega_c^\mathrm{eff}$, as discussed in Sec.~\ref{subsec:bao}, we can use a Gaussian prior $\omega_b\sim\mathcal{N}(0.02218,0.00055)$ from BBN or the compressed CMB likelihood constraining $\theta_s$ and $\omega_b$ to provide an early-Universe anchor for the BAO distances.
We consider uncalibrated SN luminosity distances from DES-Dovekie (DD), distance-ladder calibration of the SN distances (SH0ES, H0DN~\cite{H0DN:2025lyy} in the form of a Gaussian prior $H_0\sim\mathcal{N}(73.50,0.81)$) as a late-Universe geometric anchor, BAO distances (DESI), a Gaussian prior $\omega_b\sim\mathcal{N}(0.02218,0.00055)$ from BBN as an early-Universe anchor for BAO, and the $2\times2$ compressed CMB likelihood in the basis $\{\omega_b,\theta_s\}$.
The corresponding results using the Pantheon+ (PP) SN sample are presented in Appendix~\ref{appendix}.

For the $\xi\mathrm{CDM}$ model, we are able to compute $r_\mathrm{drag}$ using $\omega_c^\mathrm{eff}$, as discussed in Sec.~\ref{subsec:bao}, and therefore use BBN or the CMB as an early-Universe anchor for the BAO distances.
The discrepancy between late-time geometric anchors from the distance ladder and early-time model-dependent anchors from BBN and the CMB provides one way of characterizing the Hubble tension~\cite{Poulin:2024ken,Verde:2019ivm,DiValentino:2020zio,DiValentino:2021izs,Perivolaropoulos:2021jda,Schoneberg:2021qvd,Shah:2021onj,Abdalla:2022yfr,DiValentino:2022fjm,Kamionkowski:2022pkx,Hu:2023jqc,Verde:2023lmm,DiValentino:2024yew,CosmoVerseNetwork:2025alb,Ong:2025cwv,Cai:2026swf}.
We therefore do not attempt to combine distance-ladder-calibrated SN with early-time anchors (BBN and CMB), since the $\xi\mathrm{CDM}$ model does not alleviate the tension between these calibrations, as shown in Tab.~\ref{tab:xicdm dd table}.

\begin{table*}[h]
\renewcommand{\arraystretch}{1.3}
\centering
\setlength{\tabcolsep}{8pt}
\begin{tabular}{l | c c c c}
\hline\hline
Datasets & DD+DESI & DD+DESI+BBN & DD+DESI+H0DN & DD+DESI+CMB \\
\hline
$H_0\,[\mathrm{km}/\mathrm{s}/\mathrm{Mpc}]$ & $> 81.1$ & $65.0\pm 1.6$ & $73.52\pm 0.80$ & $67.28\pm 0.61$ \\
$\Omega_bh^2$ & $0.051^{+0.021}_{-0.010}$ & $0.02219\pm 0.00054$ & $0.0346^{+0.0028}_{-0.0032}$ & $0.02236\pm 0.00015$ \\
$\Omega_ch^2$ &  $0.204^{+0.064}_{-0.038}$ & $0.1222\pm 0.0042$ & $0.1521^{+0.0079}_{-0.0095}$ & $0.1247^{+0.0035}_{-0.0040}$ \\
$\xi_c$ &  $0.27^{+0.11}_{-0.12}$ & $0.24^{+0.10}_{-0.12}$ & $0.27^{+0.11}_{-0.12}$ & $0.101\pm 0.047$ \\
\hline
$\omega_c^\mathrm{eff}$ &  $0.157^{+0.047}_{-0.031}$ & $0.0976^{+0.0095}_{-0.011}$ & $0.1180\pm 0.0089$ & $0.1133\pm 0.0019$ \\
$L_2\,[10^4\,\mathrm{Mpc}]$ & $< 4.24$ & $ < 6.45 $ & $ < 4.92 $ & $ < 17.8 $ \\
\hline
$\Omega_m$ &  $0.347^{+0.018}_{-0.020}$ & $0.344^{+0.018}_{-0.021}$ & $0.347^{+0.018}_{-0.021}$ & $0.327^{+0.013}_{-0.015}$ \\
$r_\mathrm{drag}\,h\,[\mathrm{Mpc}] $ & $ 99.78\pm 0.78 $ & $ 99.87\pm 0.78 $ & $  99.81\pm 0.78 $ & $  100.20\pm 0.74 $ \\
$r_\mathrm{drag}\,[\mathrm{Mpc}]$ &  $119.9^{+4.0}_{-21}$ & $153.8\pm 3.3$ & $135.8\pm 1.8$ & $148.93\pm 0.55$ \\
$100\theta_s$ &  $1.052^{+0.020}_{-0.0083}$ & $1.025\pm 0.011$ & $1.0431^{+0.0089}_{-0.0081}$ & $1.04183\pm 0.00030$ \\
\hline
$\Delta\chi^2$ & $-4.85\,(2.2\sigma)$ & $-4.85\,(2.2\sigma)$ & $-1.03\,(1.0\sigma)$ & $-3.96\,(2.0\sigma)$ \\
$\ln\mathcal{B}$ & $2.77\,(2.9\sigma)$ & $0.99\,(2.0\sigma)$ & $1.35\,(2.2\sigma)$ & $-0.12$ (n/a) \\
\hline\hline
\end{tabular}
\caption{$1\sigma$ constraints on the cosmological parameters of the $\xi\mathrm{CDM}$ model for different dataset combinations using the SN dataset DES-Dovekie (DD). The upper bounds of the $L_2$ row represent the 95\% CL.}
\label{tab:xicdm dd table}
\end{table*}
\begin{figure}
    \centering
    \includegraphics[width=\linewidth]{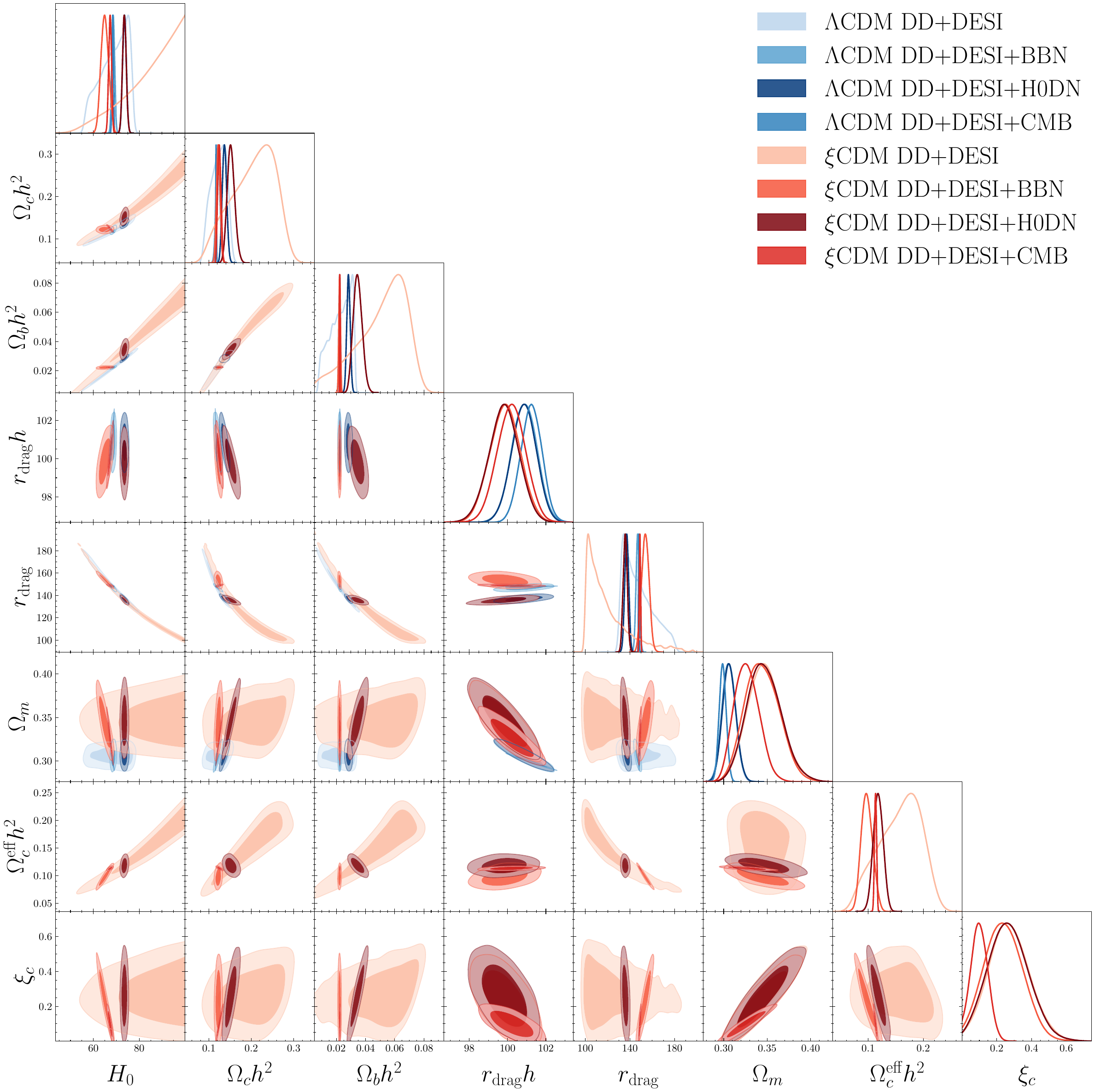}
    \caption{Posteriors of the $\Lambda\mathrm{CDM}$ and $\xi\mathrm{CDM}$ cosmologies for the DD+DESI dataset combined with the indicated anchors.}
    \label{fig:triangle anchors dm}
\end{figure}

We observe a decrease in $H_0$ when the distances are calibrated with early-Universe anchors (CMB, BBN), compared to the preferred late-time value from the distance ladder.
The $\xi\mathrm{CDM}$ model appears to prefer higher values of $\Omega_m$ compared to $\Lambda\mathrm{CDM}$.
When $\omega_b$ is constrained by BBN or the CMB, $\omega_c$ is constrained to values similar to those in $\Lambda\mathrm{CDM}$, around $0.12$.
The consequence is that the effective CDM density described by Eq.~\eqref{eq:effective cdm} is always lower than the true present-day value (since $L_2>0$).
This affects the sound horizon in Eq.~\eqref{eq:r_drag} in two ways.
First, the reduction in the effective physical CDM density results in a lower expansion rate $H(z)$ relative to $\Lambda\mathrm{CDM}$, increasing the integrand in Eq.~\eqref{eq:r_drag}.
Second, the reduction in the effective CDM density shifts the baryon drag epoch to a lower redshift, thereby decreasing the lower limit of the integral.
Both effects contribute to increasing $r_\mathrm{drag}$ relative to $\Lambda\mathrm{CDM}$ for the same present-day physical densities.

We observe this effect explicitly in the two-dimensional posterior of $r_\mathrm{drag}$ versus $\Omega_c^\mathrm{eff}h^2$ in Fig.~\ref{fig:triangle anchors dm}.
When the CMB or BBN is combined with DD+DESI, we find a negative correlation between the two parameters.
In contrast, H0DN directly constrains $H_0$ and therefore also $r_\mathrm{drag}$ through the tightly constrained combination $r_\mathrm{drag}\,h$.
The uncalibrated distance measurements alone constrain the combination $r_\mathrm{drag}\,h$ in our model.
Since the early-time anchors force $r_\mathrm{drag}$ towards larger values, this correspondingly pulls $H_0$ towards lower values, increasing the tension with the distance-ladder calibrations.

\begin{figure}
    \centering
    \includegraphics[width=0.5\linewidth]{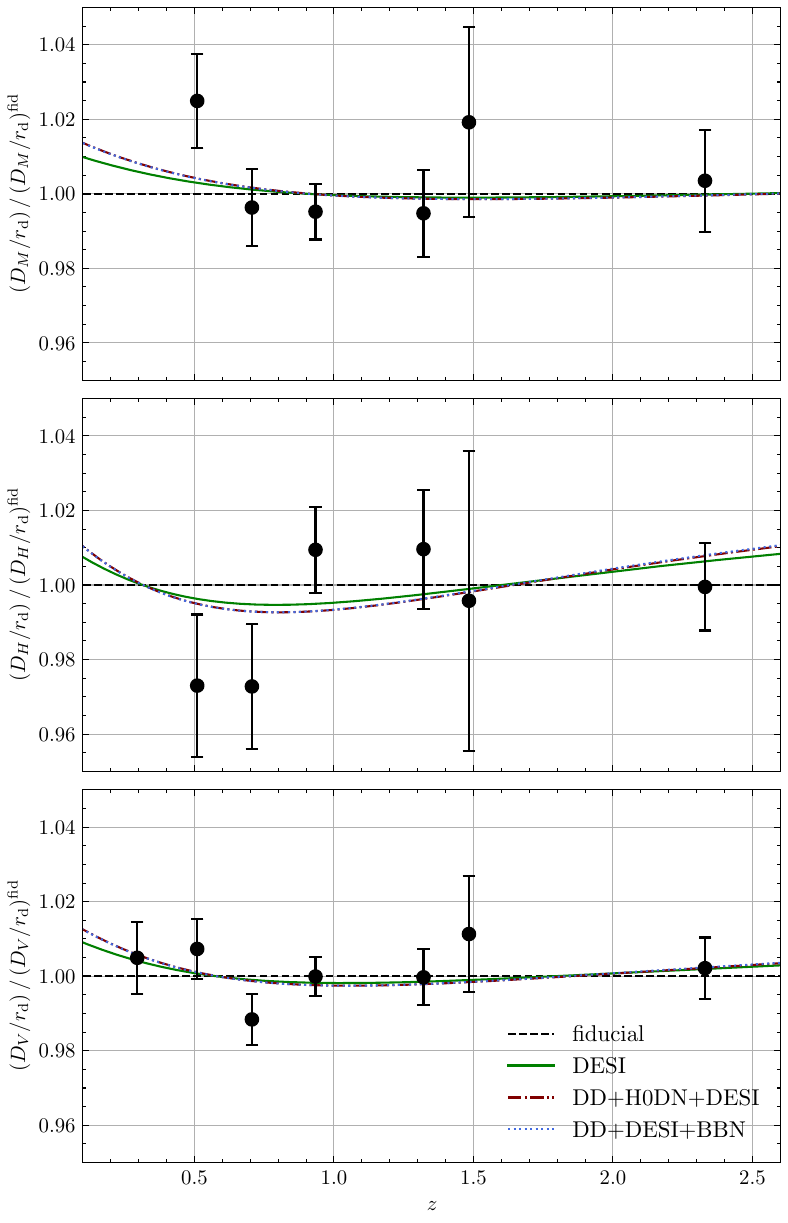}
    \caption{Residuals of the BAO distances with respect to the best-fit $\Lambda\mathrm{CDM}$ cosmology for the DESI DR2 BAO dataset.}
    \label{fig:bao residual dm}
\end{figure}

We note that the additional freedom provided by this model, through the parameter $\xi_c$, which suppresses $H(z)$ relative to $\Lambda\mathrm{CDM}$ in the past, provides a better fit to the background data than $\Lambda\mathrm{CDM}$, with a preference at the $\sim1-2\sigma$ level according to both the $\Delta\chi^2$ and $\ln\mathcal{B}$ metrics for the individual dataset combinations.
This is visually illustrated by the BAO residuals shown in Fig.~\ref{fig:bao residual dm}, where we present the distances predicted by the best-fit $\xi\mathrm{CDM}$ cosmologies for the different dataset combinations.

%
Using background data alone, as a single-parameter extension of $\Lambda\mathrm{CDM}$, $\xi\mathrm{CDM}$ performs better than $w\mathrm{CDM}$, another single-parameter extension of $\Lambda\mathrm{CDM}$ that is not significantly preferred over the standard model.
The statistical preference for $\xi\mathrm{CDM}$, expressed in terms of $\sigma$, is instead comparable to that of $w_0w_a\mathrm{CDM}$, the two-parameter extension of $\Lambda\mathrm{CDM}$ that phenomenologically describes dynamical dark energy features.

\section{Conclusions}
\label{sec:conclusions}

In this paper, we have taken a first step towards an observational assessment of the thermogravity scenario proposed in the parent Letter~\cite{Isichei:2025ssf}.
The central novelty of the framework is that the trace-free Einstein equations are supplemented by a controlled violation of energy-momentum conservation, tied to a preferred thermodynamic frame.
At the homogeneous level, this can generate late-time acceleration without a gravitating cosmological constant.
The resulting phenomenology, however, depends on how the non-conservation is distributed among the different matter species.

The minimal universal implementation provides a useful benchmark.
Using standard SN distances together with uncalibrated BAO measurements, we find that the SN data alone can be well accommodated by the model, but the inclusion of DESI BAO leads to a dramatic worsening of the fit relative to $\Lambda\mathrm{CDM}$, with $\Delta\chi^2\simeq+51$ and $\ln\mathcal{B}\simeq-26$.
The reason is the intrinsic rigidity of the universal model: acceleration is provided by universal matter creation, which forces a trade-off between the normalization of the expansion history at intermediate redshift and the amount of acceleration today.
Increasing the matter density can partially restore the intermediate-redshift expansion rate preferred by BAO, but the same change raises the present deceleration parameter, $q_0=-1+3\Omega_m/2$, thereby weakening the current acceleration.
The model is therefore pulled in incompatible directions: the SN data alone prefer $\Omega_m\simeq0.42$, while their combination with BAO shifts the preferred value to $\Omega_m\simeq0.51$, but even this shift is unable to reconcile the expansion history required by the two probes.

This conclusion should nevertheless not be over-interpreted.
A genuinely universal non-conservation law would also affect the physical systems through which cosmological observables are inferred.
Photons, baryons, clocks, redshifts, luminosities, number densities, and the calibration of standard candles and rulers would no longer be guaranteed to retain their standard operational meaning.
The construction of the Hubble diagram and the interpretation and calibration of the BAO scale would therefore have to be rederived within the theory itself.
Thus, the universal model is best regarded here as a formal benchmark: it is strongly disfavoured as a background model when the standard observational interpretation of SN and uncalibrated BAO distances is adopted, but this does not constitute a fully self-contained physical exclusion of every universal implementation, as stressed in Ref.~\cite{Isichei:2025ssf}.

The situation is different for the non-universal model in which the violation is confined to the cold dark matter sector.
In this case, baryons and photons retain their standard conservation laws, so that supernova luminosities, redshifts, BBN information, and the BAO ruler can be treated conventionally at the background level.
This makes the model both more predictive and more sharply testable.
Within the background-level analysis performed here, the $\xi\mathrm{CDM}$ model provides a systematically improved fit relative to $\Lambda\mathrm{CDM}$ for the dataset combinations considered, with improvements of up to $\Delta\chi^2=-4.85$, corresponding to about $2.2\sigma$, and positive Bayesian evidence for most combinations.
The preference for a non-zero violation parameter is typically at the $\sim1-2\sigma$ level, while the DD+DESI combination reaches a Bayesian preference of $2.9\sigma$.
The model achieves this improvement by suppressing the expansion rate relative to $\Lambda\mathrm{CDM}$ in the past.
However, it does not resolve the discrepancy between early- and late-Universe calibrations.
Once BBN or CMB information is used to calibrate the BAO ruler, the reduced effective early-time CDM density increases $r_\mathrm{drag}$; since the uncalibrated distances tightly constrain $r_\mathrm{drag}\,h$, this drives $H_0$ to lower values and therefore increases the tension with the distance-ladder calibration.
Thus, the background data provide evidence that the additional dark-sector freedom can improve the description of the expansion history, but not that this realization of thermogravity solves the Hubble tension.

These results should still be regarded as provisional.
The analysis has deliberately been restricted to the homogeneous background and to observables that can be treated without a full perturbation theory.
A definitive assessment requires the development of perturbations in the preferred-frame theory, including the CMB anisotropies, weak lensing, and structure-growth observables.
The main obstacle, and the central reason for starting with background tests, is that the choice of frame and gauge becomes physically relevant when considering fluctuations in this model.
This follows from the breaking of local Lorentz invariance and diffeomorphism invariance.
Such an analysis, although essential, is therefore considerably more model-dependent than the background treatment presented here, where the preferred frame is unambiguously specified.

The main conclusion of the present work is therefore not that thermogravity has already passed all cosmological tests, but rather that, while the universal implementation is strongly disfavoured at the background level under standard observational assumptions, the dark-sector implementation remains viable and provides a concrete and competitive target for the next stage of observational analysis.

\section{Acknowledgments} We thank Savvas Nesseris for comments on the sound horizon fitting formulae and Daniel Kessler for discussions and assistance in the \texttt{Cobaya} interface with the background theory code. 
DHL is supported by an EPSRC studentship. JM was partly supported by STFC Consolidated Grant ST/T000791/1. CvdB is supported by the Lancaster–Sheffield Consortium for Fundamental Physics under STFC grant: ST/X000621/1. EDV is supported by a Royal Society Dorothy Hodgkin Research Fellowship.
This article is based upon work from COST Action CA21136 \emph{Addressing observational tensions in cosmology with systematics and fundamental physics} (CosmoVerse), supported by COST (European Cooperation in Science and Technology).
We acknowledge IT Services at The University of Sheffield for the provision of services for High Performance Computing.

\begin{appendix} 

\section{Pantheon+ Constraints} \label{appendix}

Here we present the $1\sigma$ constraints on the UTG (Tab.~\ref{tab:utg pp table}) and $\xi\mathrm{CDM}$ (Tab.~\ref{tab:xicdm pp table}) model for the same dataset combinations as in Tab.~ \ref{tab:utg dd table} and \ref{tab:xicdm dd table} respectively, but using the Pantheon+ SN sample in place of DD and the SH0ES distance-ladder calibration of the fiducial SNIa magnitude as the late-time anchor.

\begin{table*}[h]
\renewcommand{\arraystretch}{1.3}
\centering
\setlength{\tabcolsep}{8pt}
\begin{tabular}{l | c c c c}
\hline\hline
Datasets & PP & PP+SH0ES & PP+DESI & PP+SH0ES+DESI \\
\hline
$H_0\,[\mathrm{km}/\mathrm{s}/\mathrm{Mpc}]$ & $---$ & $73.23\pm0.97$ & $---$ & $72.56^{+0.95}_{-1.10}$ \\
$\Omega_m$ & $0.414\pm0.018$ & $0.414\pm0.018$ & $0.5126\pm0.0072$ & $0.5125\pm0.0074$ \\
$r_\mathrm{drag}\,h\,[\mathrm{Mpc}]$ &  $---$ & $---$ & $95.52\pm0.57$ & $95.53\pm0.57$ \\
\hline
$\xi_c$ &  $1.071\pm0.041$ & $1.071\pm0.041$ & $0.852\pm0.015$ & $0.852\pm0.015$ \\
$L_2\,[10^3\,\mathrm{Mpc}]$ &  $4.28^{+0.60}_{-2.00}$ & $3.83^{+0.15}_{-0.17}$ & $5.32^{+0.70}_{-2.00}$ & $4.85\pm0.11$ \\
$t_0\,[10^3\,\mathrm{Mpc}]$ &  $4.74^{+0.60}_{-2.00}$ & $4.25\pm0.10$ & $4.22^{+0.50}_{-1.00}$ & $3.85\pm0.06$ \\
$r_\mathrm{drag}\,[\mathrm{Mpc}]$ &  $153^{+20}_{-50}$ & $136.5^{+7.6}_{-10}$ & $144^{+20}_{-50}$ & $131.7\pm2.0$ \\
\hline
$\Delta\chi^2$ & $1.30$ & $1.89$ & $46.7$ & $47.5$ \\
$\ln\mathcal{B}$ & $-0.66$ & $-0.97$ & $-23.48$ & $-23.90$\\
\hline\hline
\end{tabular}
\caption{$1\sigma$ constraints on the cosmological parameters of the UTG model for different dataset combinations using the SN dataset Pantheon+ (PP) sample.}
\label{tab:utg pp table}
\end{table*}
\begin{table*}[h]
\renewcommand{\arraystretch}{1.3}
\centering
\setlength{\tabcolsep}{8pt}
\begin{tabular}{l | c c c c}
\hline\hline
Datasets & PP+DESI & PP+DESI+BBN & PP+DESI+SH0ES & CMB+PP+DESI \\
\hline
$H_0\,[\mathrm{km}/\mathrm{s}/\mathrm{Mpc}]$ & $> 80.3$ & $65.0\pm 1.7$ & $73.6\pm 1.0$ &  $67.35^{+0.70}_{-0.63}$\\
$\Omega_bh^2$ &  $0.050^{+0.022}_{-0.011}$ & $0.02218\pm 0.00056$ & $0.0345^{+0.0028}_{-0.0032}$ &  $0.02235\pm 0.00015$\\
$\Omega_ch^2$ &  $0.201^{+0.065}_{-0.040}$ & $0.1221\pm 0.0043$ & $0.1519^{+0.0086}_{-0.0097}$ &  $0.1243^{+0.0036}_{-0.0045}$\\
$\xi_c$ & $0.27^{+0.11}_{-0.13}$ & $0.24\pm 0.11$ & $0.26^{+0.11}_{-0.13}$ &  $0.097^{+0.047}_{-0.055}$\\
\hline
$\omega_c^\mathrm{eff}$ &  $0.156^{+0.048}_{-0.033}$ & $0.0979^{+0.0094}_{-0.011}$ & $0.1187\pm 0.0091$ &  $0.1133^{+0.0020}_{-0.0018}$\\
$L_2\,[10^4\,\mathrm{Mpc}]$ & $ < 4.87$ & $ < 8.19 $ & $ < 5.63 $ & $ < 22.1$ \\
\hline
$\Omega_m$ &  $0.346^{+0.019}_{-0.023}$ & $0.343^{+0.020}_{-0.022}$ & $0.346^{+0.019}_{-0.023}$ & $0.325^{+0.013}_{-0.017}$\\
$r_\mathrm{drag}\,h\,[\mathrm{Mpc}]$ & $  99.82\pm 0.85 $ & $  99.90\pm 0.85 $ & $  99.83\pm 0.84 $ & $  100.30\pm 0.79 $\\
$r_\mathrm{drag}\,[\mathrm{Mpc}]$ & $120.6^{+4.8}_{-21}$ & $153.7\pm 3.2$ & $135.7\pm 2.1$ &  $148.92\pm 0.55$\\
$100\theta_s$ &  $1.052^{+0.021}_{-0.0089}$ & $1.025\pm 0.011$ & $1.0436^{+0.0090}_{-0.0080}$ &  $1.04184\pm 0.00030$\\
\hline
$\Delta\chi^2$ & $-4.05\,(2.4\sigma)$ & $-4.05\,(2.0\sigma)$ & $-4.01\,(2.0\sigma)$ & $-2.93\,(1.7\sigma)$\\
$\ln\mathcal{B}$ & $2.40\,(2.7\sigma)$ & $0.66\,(1.8\sigma)$ & $0.99\,(2.0\sigma)$ & $-0.58$ (n/a)\\
\hline\hline
\end{tabular}
\caption{$1\sigma$ constraints on the cosmological parameters of the $\xi\mathrm{CDM}$ model for different dataset combinations using the Pantheon+ (PP) SN sample. The upper bounds of the $L_2$ row represent the 95\% CL.}
\label{tab:xicdm pp table}
\end{table*}

\end{appendix}

\bibliographystyle{apsrev4-1}
\bibliography{bibliography}

\end{document}